# The Gap Test – Effects of Crack Parallel Compression on Fracture in Carbon Fiber Composites

**Jeremy Brockmann**
Graduate Researcher
Master of Science Program in the Department of
Aeronautics & Astronautics
University of Washington
Seattle, Washington 98195

**Marco Salviato** *
Associate professor of Aeronautics & Astronautics
Department of Aeronautics & Astronautics
University of Washington
Seattle, Washington 98195
Email: salviato@aa.washington.edu

### ABSTRACT

    This paper explores the global Mode I fracture energy of a carbon fiber composite subject to a biaxial stress state at a crack tip, specifically in which one stress component is compressive and parallel to the crack. Based on an experimental technique previously coined as *The Gap Test* and Bažant's Type II Size Effect Law, it is found that there is a monotonic decrease in the Mode I fracture energy as the crack parallel compressive stress increases. Compared to the nominal value of fracture energy, where no crack parallel compression is applied, the fracture energy is observed to decrease by up to 37% for a compressive stress equal to 44% of the compressive failure limit of the composite. This weakening effect is attributed to splitting cracks that are induced at the crack tip due to the crack parallel compression, which are identified via crack tip photomicroscopy. This is a novel result that challenges the century old hypothesis of fracture energy being a constant material property and further, shows for the first time that crack parallel compression leads to a composite structure being dangerously weaker than expected.
    The experimental campaign is also buttressed with a computational campaign that provides a framework capable of capturing the effects of crack parallel compression. Through the use of the crack band model, which correctly characterizes the fracture process zone tensorially, coupled with a fully tensorial damage law, the simulated results provide satisfactory agreement with the experimental data. Conversely, when a reduced tensorial damage law defines the crack band it is shown that the structural strength and fracture energy are dangerously overpredicted. This emphasizes the importance of using a crack band model coupled with a fully tensorial damage law to accurately predict fracture in composites.

*Address all correspondence to this author.

**Keywords: A.** Polymer-matrix composites (PMCs), **B.** Fracture Toughness, **C.** Finite Element Analysis (FEA), the Gap Test

# 1 Introduction

Carbon fiber composites have experienced an exponential increase in their structural applications over the recent decades. They are found ubiquitously today in the aerospace [1], energy [2], automotive [3], recreational [4], and civil [5] industries. A primary driver for this phenomenon is composites' superior strength-to-weight ratio when compared to traditional metals such as steel, aluminum, and titanium. Furthermore, composites offer greater corrosion resistance, impact energy adsorption, and fatigue life while their mechanical properties are also tailorable [6].

In the context of the increasing success of composites, the present study is intended to contribute to engineers' understanding of these materials and how they may be used to better meet our societal needs. Specifically, designing against fracture in engineering structures which is a formidable problem that has been studied by engineers for over a century [7]. Current advancements in fracture theory, of which many have unfortunately been learned through fatal accidents – airline crashes [8], civil structure failures, and pressure vessel explosions [9] are examples, have yet to lead to a reliable numerical or theoretical framework for the prediction of progressive failure (and strength) in composites that is driven by the unique failure mechanisms at multiple length scales. Clearly a better fracture theory is imperative to guarantee safety and reliability in the increasingly complex composite structures of today.

The most ubiquitous fracture theory is Linear Elastic Fracture Mechanics (LEFM), whose concepts are directly applicable only to materials that behave in an ideally brittle manner. Additional foundational assumptions of LEFM are that any damage or plasticity that occurs as a crack propagates are negligible in size compared to the structure size and the overall response of the structure is linear elastic [10]. Stemming from this, cracks in LEFM are modeled as a line, mathematically having zero width and a crack tip radius equal to zero. Lastly, LEFM leverages a single parameter description of fracture where most commonly the parameters critical fracture energy $G_c$ or the critical stress intensity factor $K_c$ are used to predict fracture and are assumed to be constant material properties.

The assumptions defining LEFM are often too restrictive to apply to many modern engineering materials. Specifically, when a material deviates from ideally brittle behavior LEFM is not applicable. This phenomenon was discovered in concrete and later extended experimentally to shale, fiber composites, sea ice, rocks, ceramics, bone, and many more (for a concise review, please refer to [11] and [12]). In these circumstances the material may exhibit a quasibrittle behavior and hence a new fracture theory, Quasibrittle Fracture Mechanics (QBFM), is needed. The present paper focuses on QBFM as its assumptions are the most appropriate for fracture in composites [11, 13, 14, 15, 12, 16, 17, 18, 19, 20].

A key characteristic of QBFM is the size effect. The size effect is understood as the structural strength dependence on structure size. This is a departure from classical strength predictions such as plastic-limit or strength of materials analysis, which predict no size effect. On a very small scale quasibrittle materials display pseudo-plastic behavior, on a medium scale they display quasibrittleness, and on large scales they display closer to ideal brittleness. This transition in material behavior implies the presence of a non-negligible characteristic length of the material. This is taken to be the size of the Fracture Process Zone (FPZ). The FPZ is a large, non-linear region of damage that surrounds the crack tip during fracture in a quasibrittle media. LEFM theory does not account for the effects of a FPZ nor does it account for the transitional behavior of a quasibrittle material. To define the size effect in quasibrittle materials Bažant's Type II Size Effect Law (SEL) must be used [21]:

$$\sigma_N = \sigma_o(1 + D/D_o)^{-1/2} \tag{1}$$

where D denotes a characteristic length, $\sigma_N$ denotes structural strength, $\sigma_o$ is an arbitrary measure of material strength, and $D_o$ is a reference geometric size.

## *1.1 Gap Test Motivation and Goal*

Many existing fracture theories rest upon a century old hypothesis that a material's critical fracture energy is constant [7, 10], being defined as a material's resistance to crack propagation. One attributor to constant fracture energy is the modeling of a crack as a line, as done in LEFM. When line crack models are used the crack may only be opened via 3 modes – in-plane crack opening normal stresses (Mode I), in-plane shear stresses (Mode II), and out-of-plane shear stresses (Mode III). If a planar biaxial stress state is then considered, in which one stress component is perpendicular to the crack plane and one is parallel, the parallel stress has no ability to open the crack. The biaxial stress state is thus equivalent to a uniaxial stress state, at least regarding fracture, when a line crack model is used. This is illustrated in Figure 1 where any damage or plasticity at the crack tip is considered negligibly small according to LEFM. So, because the x-direction line crack in Figure 1 has no change on $\sigma_{xx}$, the biaxial stress state has no impact on the structure's ability to resist fracture. Thus, it is concluded that the fracture energy is constant under the shown uni-axial and bi-axial stress states.

      In the presence of a finite width FPZ, as is the case for quasibrittle media such as composites, the assumption of a line crack breaks down. Consider again the stress state shown in Figure 1 but with the inclusion of a FPZ that is non-negligible compared to the structure size. This is shown in Figure 2. Now, because the FPZ has a non-negligible width in the x direction, the stress $\sigma_{xx}$ that is parallel to the crack plane will have an influence on the crack opening behavior meaning the stress state is not equivalent to that of a uni-axial stress state, as was the case for the LEFM assumption in Figure 1. This has 2 novel consequences. First, it implies that a crack may propagate by way of a fracture mode that is not defined by Mode I, II, or III. Second, because the crack parallel stress $\sigma_{xx}$ will influence the structure's ability to resist fracture, the fracture energy is likely to be non-constant. This conclusion posits an intriguing question that governs the present study: *does the presence of a crack parallel stress influence the constancy of fracture energy in composites?*

## *1.2 Review of the Gap Test*

Due to the novelty of the ideas posed in the previous section the existing academic literature is absent in relevant studies on composite materials but a trailblazing study was performed by Bažant and co-workers at Northwestern University in 2020 where an equivalent experiment was performed on concrete specimens [22, 23]. Concrete is known to behave in a quasibrittle manner and hence these studies are tantamount to a composite study. Much inspiration in this study is drawn from the efforts and results of the concrete studies at Northwestern.

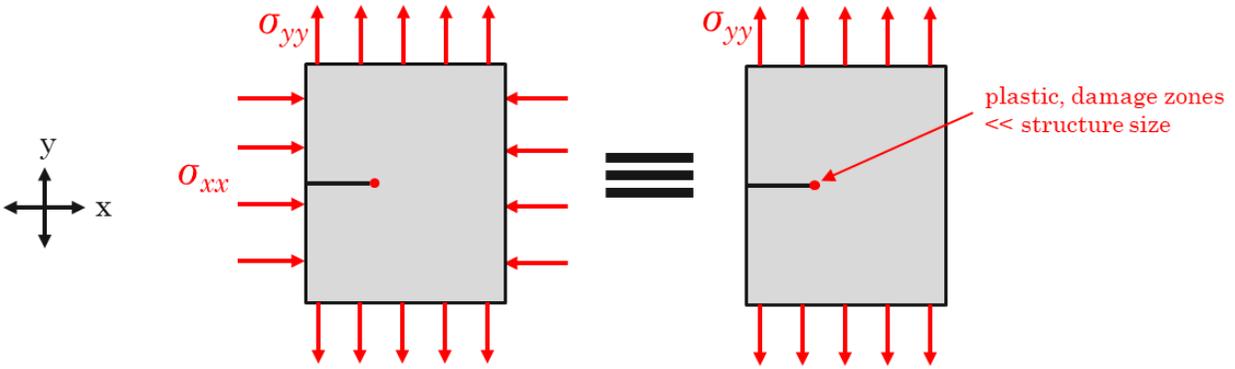

Figure 1 Equivalence of a biaxial and uniaxial stress state, regarding fracture behavior, according to the assumptions of LEFM.

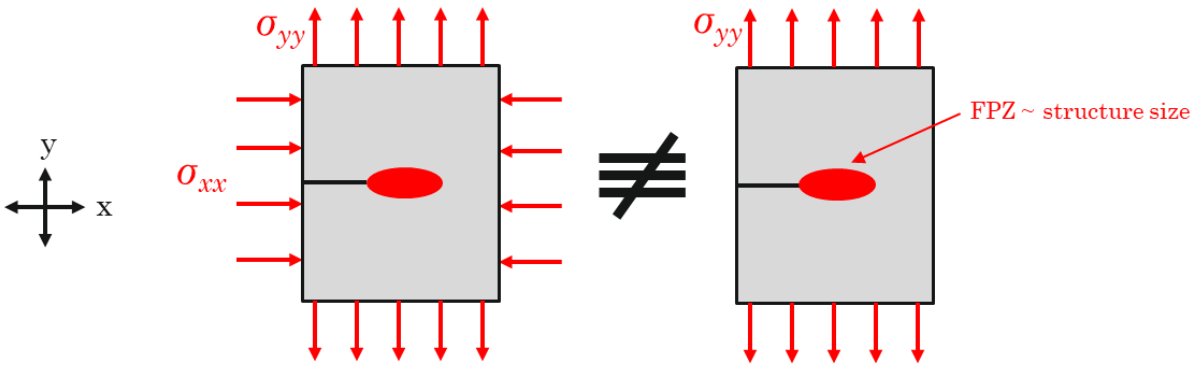

Figure 2 Influence of a finite width FPZ on a quasibrittle material under a biaxial stress state.

    The studies performed at Northwestern University investigated the same stress state as the one proposed in Figure 2, a crack parallel compression with a bending moment whose normal stresses open the crack via Mode I fracture. To accomplish this, they employed a simple modification to the standard 3PB test where a polymer pad with a perfectly plastic yield plateau is used to generate a crack parallel compression. The test specimen is placed atop the pads which are designed in such a way that they yield completely prior to engaging rigid rollers that induce a bending moment [22]. A schematic of this test procedure is shown in Figure 3, which these researchers dubbed as *The Gap Test*.

    The results from the Northwestern studies suggest that concrete exhibits a non-constant fracture energy when subject to a crack parallel compression. Specifically, they observed that, depending on the level of compression applied, the fracture energy can increase up to 1.8 times the nominal Mode I fracture energy or reduce it to almost zero as the

compression level approaches the compressive strength of the structure [22]. A similar trend is observed for the $c_f$ value, which is a measure of the FPZ size[1].

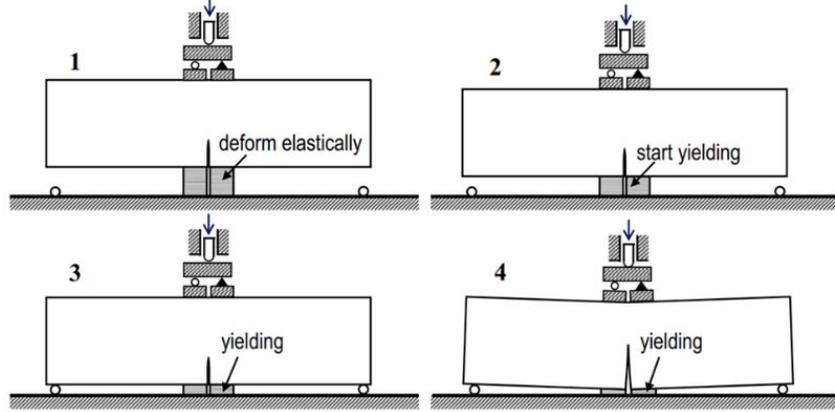

Figure 3 Experimental setup used on concrete specimens by [22, 23].

## 2 Composite Gap Test: Experiment Design and Manufacturing

### *2.1 Experimental Considerations*

This project explores the fracture behavior of composites that are subject to a biaxial stress state in which one stress component is parallel to a crack. It is believed that the fracture energy will fluctuate based on the level of biaxiality. Due to the novelty of the Gap Test a crack parallel compressive stress is chosen for experimentation on composites, like that of [22, 23]. Further, the experimental setup used in the concrete studies at Northwestern, shown in Figure 3, will serve as the backbone of the setup used to test on composites.

### *2.1.2 Buckling Analysis*
In adapting the Gap Test for composites, a key difference between concrete and composites must be considered – the characteristic thickness. Concrete specimens are typically thick, brick-like structures whereas composites are thin, shell-like structures that are susceptible to buckling. Hence, the following simple design criterion is imposed to ensure acceptable experimental results:

$$\eta P_{fracture} < P_{buckle} \qquad (2)$$

where $P$ denotes a load and $\eta$ is a factor-of-safety-like value. In the present study $\eta = 2$ was used. Both the fracture and buckling analyses are dependent upon the composite material system's mechanical properties and the laminate stacking sequence. The present study uses a symmetric cross-ply laminate (0° and 90° layers only) and a Toray T800H/3900-2 material system whose material properties are shown in Table 1 and Table 2. The analytical details of satisfying equation (2) are given in Appendix A which results in a thickness of t = 7.6 mm and indicates a 38-ply laminate.

---

[1] ***Comment on T-Stress.*** T-stress is typically used as a measure of plastic constraint at the crack tip in a ductile metal and it may be measured experimentally or computationally [44]. Depending on the geometry of the finite plate (or structure) the T-stress term may act like a crack parallel stress which has been alluded to in the conceptualization of the Gap Test. Thus, it may appear that crack parallel stress, disguised as T-stress, is a known concept that has been researched with the origination of LEFM and consequently the Gap Test could yield trivial results or, the conclusions to draw from it have already been made. This concern is not legitimate for two reasons. First, the effects from a T-stress are small in comparison to the crack parallel stresses applied in the Gap Test, never approaching the failure limit of the material. Second, and more forcefully, the concept of T-stress was derived for an isotropic, linear-elastic, and metallic material meaning it is not transplantable to quasibrittle materials in which the physics and governing material behavior are different [22].

| Elastic Properties (GPa) | |
|---|---|
| $E_{11}$ | 152.4 |
| $E_{22}$ | 9.205 |
| $G_{12}$ | 4.275 |
| $v_{12}$ | 0.35 |

ply thickness = 0.2 mm

Table 1 Mechanical properties of a Toray T800H/3900-2 uni-directional lamina per [24].

| Strength Properties (MPa) | |
|---|---|
| Longitudinal tensile strength, $X_T$ | 2089.0 |
| Longitudinal compression strength, $X_c$ | 1482.0 |
| Transverse tensile strength, $Y_T$ | 79.29 |
| Transverse compression strength, $Y_C$ | 231.0 |
| Shear strength, S | 132.8 |

Table 2 Strength properties of a Toray T800H/3900-2 uni-directional lamina per [24].

### 2.1.3 Specimen Geometry

The composite Gap Test will also leverage size effect testing, meaning equation (1) is the primary means of analysis. This allows for evaluation of fracture properties as a function of size [25], a governing characteristic of QBFM. Test specimens that are scaled in-plane by values of 1:2:4 were used while the thickness is held constant. Predicated on the buckling analysis the dimensions in Figure 4 are determined and the specimen geometry used is given in Table 3 and pictured in Figure 5. Also, using pads in the Gap Test that have a near perfectly plastic yield plateau is imperative to the test's success. In [22, 23] polypropylene pads are used. Polypropylene is a cheap and readily available material in the location in which the composite Gap Tests were performed and hence polypropylene is also used. Lastly, the Gap Test requires a single edge notched bend (SENB) specimen that is cracked to a pre-determined length. In metallic materials this is traditionally accomplished via fatigue pre-cracking [26], although fatigue pre-cracking is much more difficult in composites and can prove to be prohibitive timewise. Cracks are thus installed manually with a thin kerf blade.

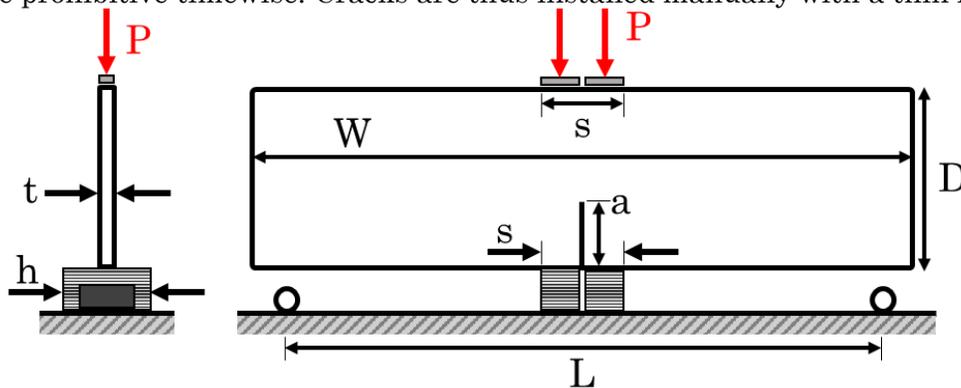

Figure 4 Gap Test schematic show test specimen geometry.

| Panel Type | Panel Width, D (mm) | Roller Span, L (mm) | Top Span, s (mm) | Panel Length, W (mm) | Crack Length, a (mm) |
|---|---|---|---|---|---|
| Small | 19.05 | 76.2 | 12.7 | 83.82 | 9.525 |
| Medium | 38.1 | 152.4 | 25.4 | 165 | 19.05 |
| Large | 76.2 | 304.8 | 50.8 | 335.28 | 38.1 |

*thickness, t = 7.6 mm for all panels, pad length h is fixed at h = 25.4 or 38.1 mm depending on the compression level

Table 3 Test panel dimensions based on Figure 4 geometry.

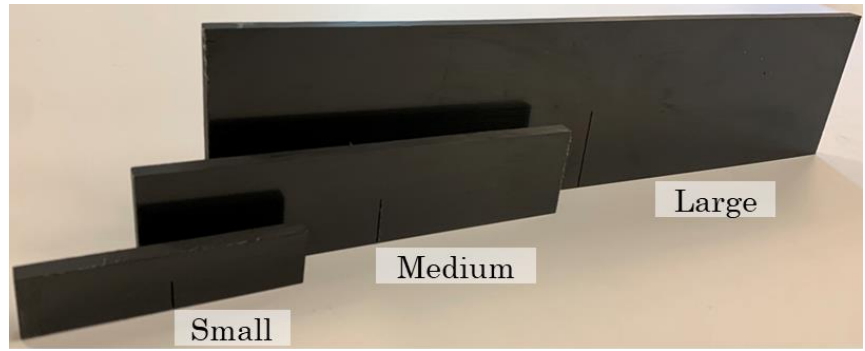

Figure 5 Size effect composite test specimens that are scaled in plane while the thickness is held constant.

### 2.1.4 Specimen Preparation

The composite material used is a pre-impregnated material that is stored in a freezer at -17 °C. The composite is wound around a cardboard cylinder and is stored in the freezer in such a way that the carbon fibers do not warp or break. It is also emphasized that the composite material is stored in an air-tight bag in the freezer, any moisture intrusion into the matrix may result in porosity or incomplete curing in the final cured composite part. Prior to using the composite material, it is removed from the freezer and allowed to thaw such that it will be pliable enough to work with. After thawing, it is rolled out on a ply cutter table that drags a sharp knife across the pre-preg composite to cut it to the exact dimensions of each ply. The ply cutter table pulls vacuum on the pre-preg during the cutting operation to ensure a clean and precise cut is achieved.

Next, the composite plies are stacked by hand atop a flat aluminum plate to form the desired shape, size, and stacking sequence of the cured composite part. During layup the first ply down, every fourth ply after, and the last ply down are debulked for 10 minutes at a minimum vacuum pressure of 26 inHg. The layup schedule is a $[(0/90)_{9.5}]_s$ laminate that takes approximately 6 hours to layup entirely.

After the layup is complete, the cure bag is applied to the part. A quality check of the cure bag is performed via a drop test, in which the cure bag is installed, brought up to full vacuum pressure, the vacuum pump is turned off, and the cure bag is required to maintain a certain vacuum pressure for a duration of time. For the present manufacturing process a drop test requirement of at least 26 inHg, with a drop no larger than 1 inHg, for 10 minutes is imposed. After the cure bag has passed the drop test it is cured via a single dwell (180 minutes), high temperature (177°C), and high-pressure (655 kPa) autoclave cure. The total cure time including controlled temperature ramps at 0.8°C/minute is ~10 hours. Images from the composite manufacturing process are shown in Figure 6.

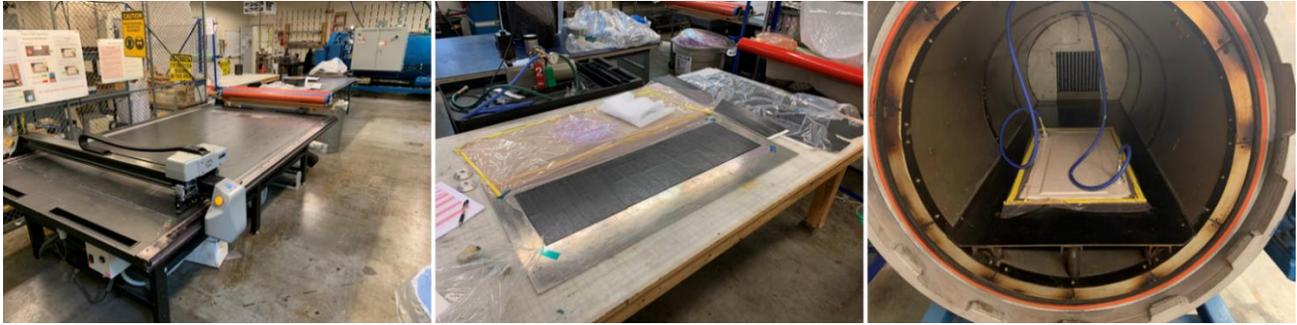

Figure 6 (left) CNC ply cutter cutting pre-preg composite into the desired size, shape, and quantity of plies (center) in-process pre-preg layup (right) completed layup loaded in an autoclave for cure.

The composite manufacturing process outlined is used to fabricate a single master panel that multiple test specimens are cut out of to the dimensions defined in Table 3. The tests specimens are cut a minimum of 25.4 mm (1") from the edge of the master panel to ensure that no uneven ply terminations are included in the test specimens. This partitioning presents the opportunity to inspect the cross section for porosity via photo-microscopy (PMG), cross section images are shown in Figure 7 for different magnifications. No visible porosity is identified which indicates a high-quality composite panel and from which, there are no concerns that quality defects will undermine the experimental results.

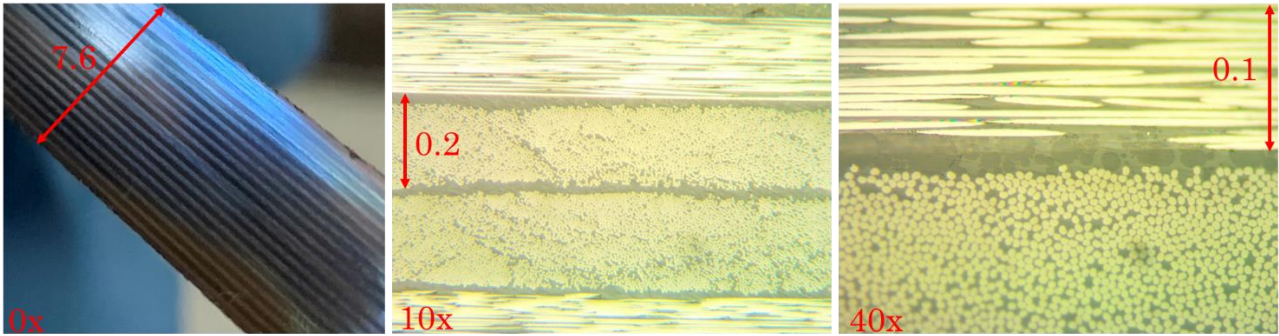

Figure 7 PMG images of the manufactured composite panel's cross section at 0x, 10x, and 40x magnification, shown values are in units of mm.

The ensuing step in test specimen preparation is installation of the cracks, for which a standard procedure for pre-cracking composites does not exist. Metallics are typically pre-cracked through fatigue loading at a stress amplitude that is much lower that the material's failure limit [26] which, is an effective technique though it is expensive timewise and requires dedicated fixturing. Furthermore, the anisotropic behavior and heterogenous composition of composites make it unlikely that a straight, self-similar crack will be achieved through fatigue pre-cracking. Cracks for the present study are installed by hand using the thinnest kerf blade that could be found for purchase, a 0.508 mm kerf blade from Zona Tool. To guarantee a straight crack is installed a mitre box is used to guide the blade to the desired crack length, avoiding off angled cracks that would lead to mixed mode loading and consequently impact the specimen's structural strength [14].

*2.1.5 Test setup*

As this study focuses on an avant-garde experimental technique all of the necessary test fixturing is not likely to be available, which was the case for the authors of this paper. Ultimately three different fixture designs were machined out of 1018 Low-Carbon Steel blanks and tested to achieve the desired loading conditions depicted in Figure 3. In the final design iteration lateral stability columns with a bolted constraint were required to prevent the test specimens from toppling and ensure they stay within the plane in which the compressive load is applied. The bolt is fastened only finger tight and consequently any contribution of the stability columns to the test specimen's fracture behavior is assumed to be negligible. This fixture is shown in Figure 8.

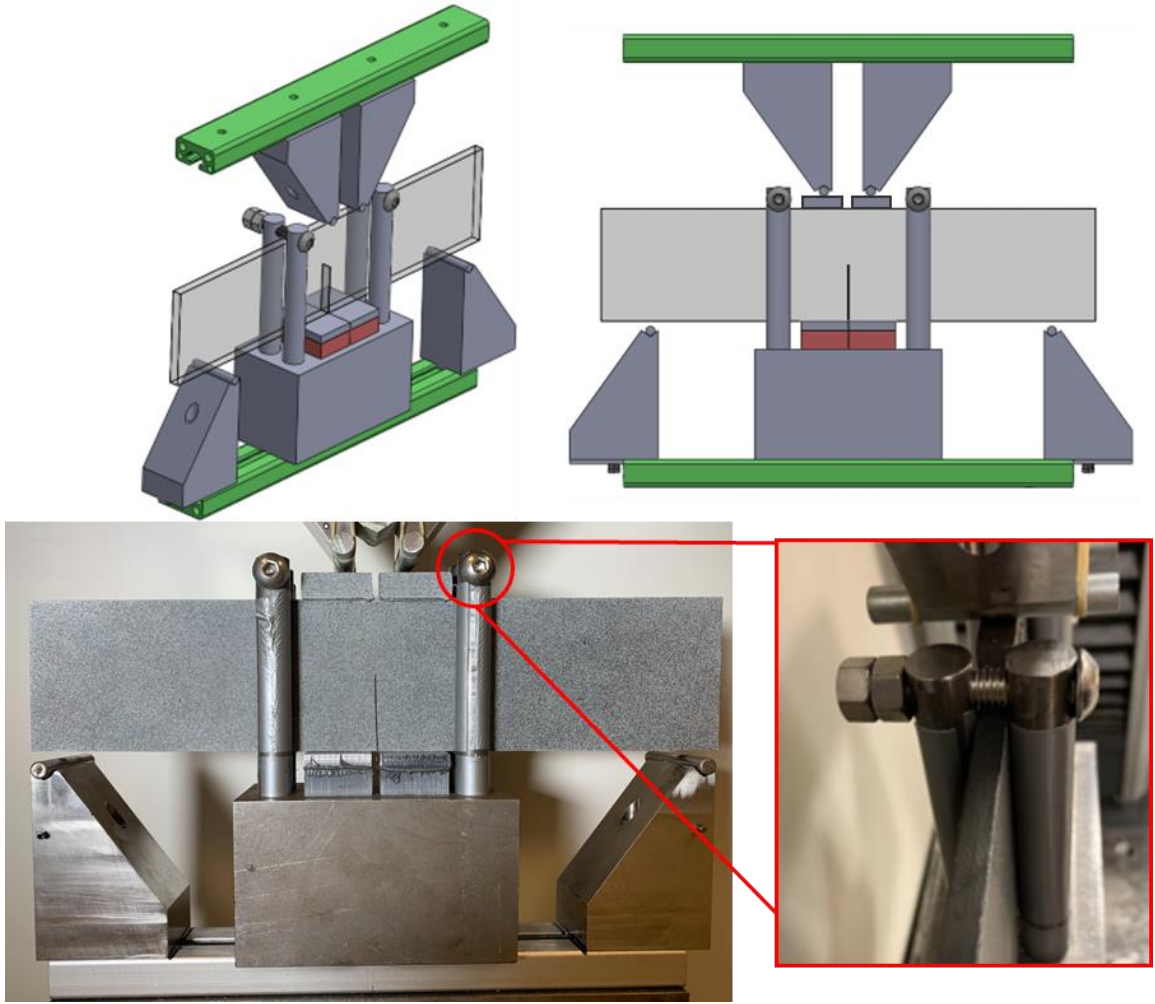

Figure 8 Final Gap Test design used to capture all experimental data.

## 3 Composite Gap Test: Results & Analysis

The Gap Tests herein are executed on an Instron 5585H load frame at a constant displacement of 1.27 mm/min, where force is measured on the load frame transducer and displacement is measured both by this transducer and optically via digital image correlation. The general setup is shown in Figure 9.

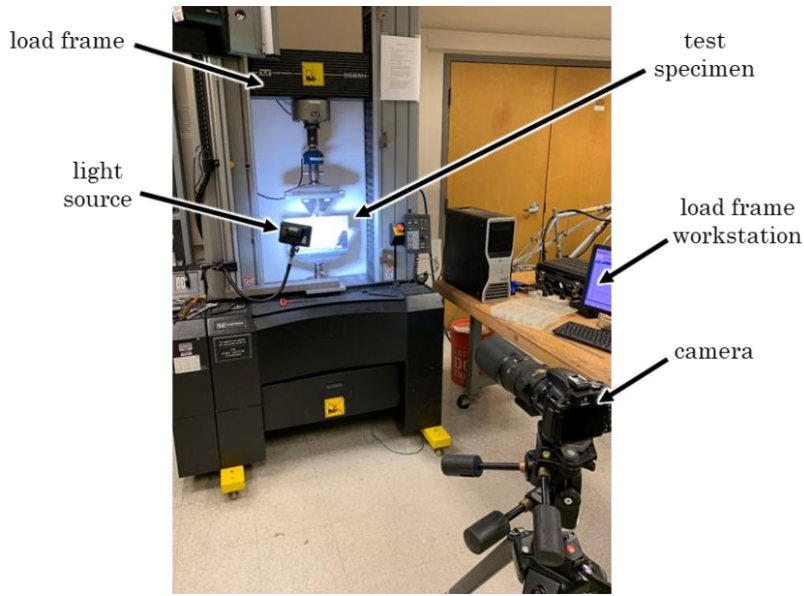

Figure 9 Experimental setup used to acquire the Gap Test data.

A combined force displacement plot for all 3 panels sizes, subsequently referred to as small, medium, and large per Table 3, of the control tests is shown in Figure 10. Notice that for all test sizes the force displacement response is linear up to the peak load, indicative of brittle behavior as there is no warning of failure or crack propagation and implies a limited effect of the nonlinear stresses within the FPZ. The control tests also display progressive damage, deduced by the saw-tooth behavior of the force displacement response immediately prior to and or after the peak load. A stable post-peak regime is also seen in all the control tests which corresponds to stable crack propagation during the fracture test. Similar stable post-peak regimes were reported in [27] where a novel fixture for Compact Tension (CT) tests was developed to induce stable strain softening. For analysis via the Bažant SEL the peak load is the most critical experimental value which is tabulated in Table 4.

The same test procedure is performed on all the test specimen sizes but now including the crack parallel compression which is achieved via the design in Figure 8. Two levels of compression are tested, $\xi = 0.29$ and $\xi = 0.44$, where $\xi$ is the crack parallel stress normalized by the compressive strength of the laminate. By this definition $\xi = 0$ indicates no crack parallel compression and $\xi = 1$ indicates the material is at its compressive failure limit. The compressive strength of the laminate $\sigma_c$ is estimated analytically via the Tsai-Wu failure criterion [28], yielding a compressive strength of $\sigma_c = 701.36$ MPa.

| | $\xi = 0$ | | |
|---|---|---|---|
| Size | Quantity | Avg. Peak Load (N) | CoV* |
| Small | 4 | 7167.2 | 0.0834 |
| Medium | 4 | 9924.1 | 0.0702 |
| Large | 3 | 13631.2 | 0.1001 |

*CoV = Coefficient of Variation

Table 4 Peak load values for the Gap Test with no crack parallel compression ($\xi = 0$).

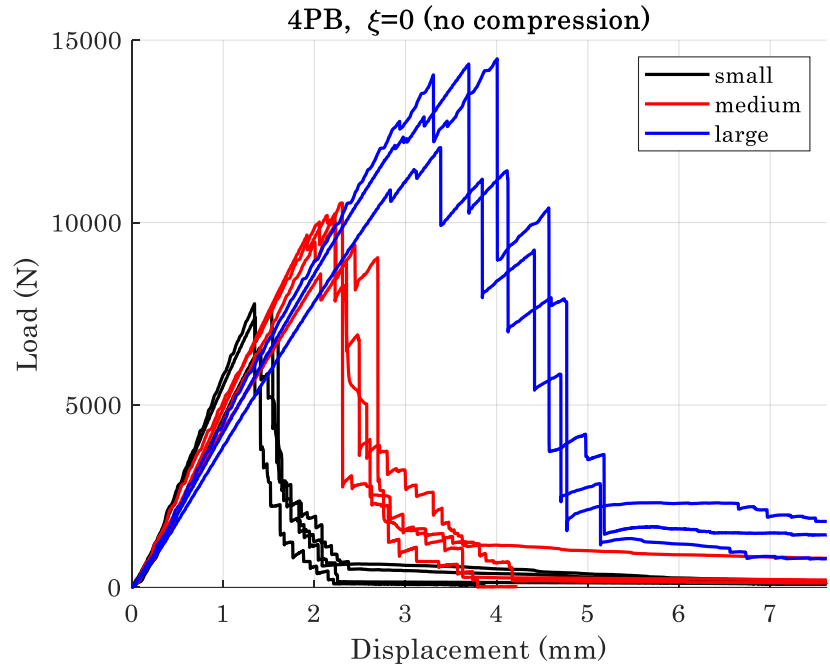

Figure 10 Force displacement results for the Gap Test with no crack parallel compression.

The force displacement curves for the ξ = 0.29 and ξ = 0.44 tests are notably different in shape than the curves for ξ = 0 due to the presence of the compressive stress but, the four distinct stress states are easily identifiable within these curves as seen in the representative plot in Figure 11. In contrast with the ξ = 0 data, the ξ = 0.29 and ξ = 0.44 tests also tend to display an unstable post-peak behavior (aka dynamic fracture), noted by the sudden drop of the force displacement plot which is characteristic of unstable crack propagation and or a catastrophic failure. This is an early indicator that different, or perhaps more, damage mechanisms are occurring in the FPZ due to the presence of the crack parallel compression.

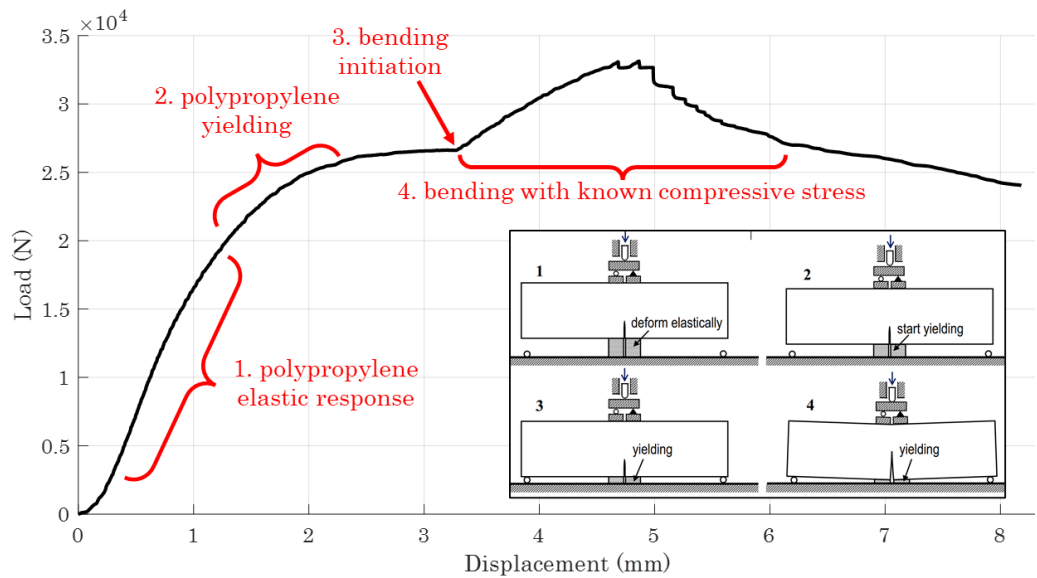

Figure 11 Example force displacement plot for the Gap Test with crack parallel compression decomposed into the 4 distinct, statically determinate stress states.

The peak loads of the bending regimes for the ξ = 0.29 and ξ = 0.44 tests are identified and shown graphically in Figure 12. Comparison of these values indicates a decrease in the peak load and structural strength as ξ is increased which correlates to a reduction of the fracture energy, contradicting the LEFM assumption that the critical fracture energy is a constant material property. This is further supporting evidence that the finite width FPZ in composites is impacted by a crack parallel compression.

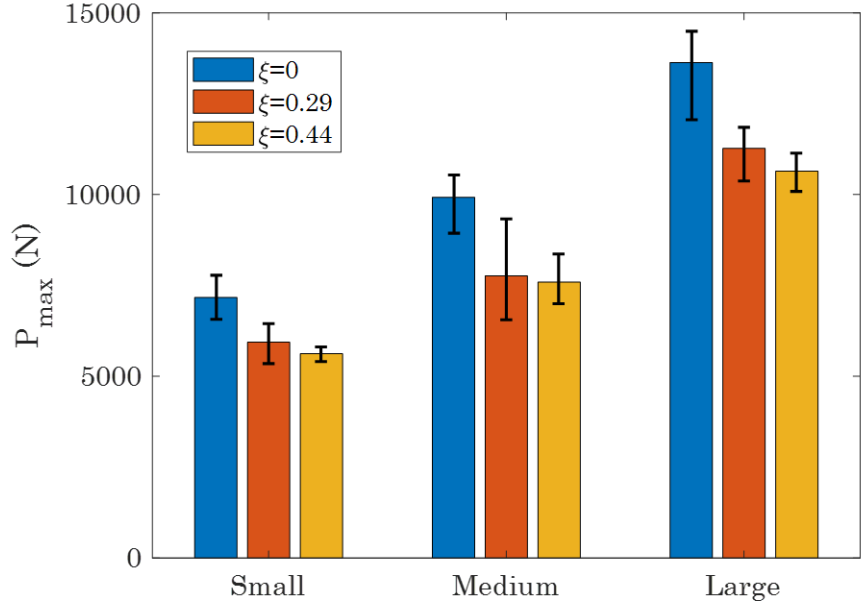

Figure 12 Measured peak loads for ξ = 0, 0.29, and 0.44. The x-axis label refers to panel type i.e., the geometrically scaled in-plane dimensions.

### 3.1 Size Effect Analysis

The Bažant SEL is repeated in the following for convenience:

$$\sigma_N = \sigma_o(1 + D/D_o)^{-1/2} \quad (3)$$

$$\sigma_o = \sqrt{EG_f/c_f g'(\alpha_o)} \quad (4)$$

$$D_o = c_f g'(\alpha_o)/g(\alpha_o) \quad (5)$$

where $G_f$ is the fracture energy, $c_f$ is a measure of the FPZ size, $D$ is a characteristic structure size, and $\sigma_N$ is the structural strength. Manipulation of equation (3) leads to the linear expression shown below,

$$\begin{gathered} \sigma_N = \sigma_o(1 + D/D_o)^{-1/2} \\ \downarrow \\ 1/\sigma_N^2 = 1/\sigma_o^2 + D/D_o\sigma_o^2 \\ \downarrow \\ \text{let } Y = 1/\sigma_N^2, C = 1/\sigma_o^2, A = 1/D_o\sigma_o^2 \\ \downarrow \\ \therefore Y = C + AD \end{gathered} \quad (6)$$

which is used to fit the experimental data using a linear regression plot of $1/\sigma_N^2$ versus D. From the slope, A, and intercept, C, of this fit the fracture properties $G_f$ and $c_f$ may be determined for each data set ξ = 0, 0.29, and 0.44 using equations (7) and (8). That is,

$$G_f = \frac{g(\alpha_o)}{EA} \tag{7}$$

$$c_f = \frac{Cg(\alpha_o)}{Ag'(\alpha_o)} \tag{8}$$

To utilize equations (6), (7), and (8) the peak load values reported previously are converted to structural strength via the application of beam theory to the 4PB loading shown in Figure 4. This gives,

$$\sigma_N = \frac{3P(L-s)}{2D^2 t} \tag{9}$$

The aforementioned equations also require the effective modulus of elasticity of the laminate, from classical lamination theory this is calculated to be E = 83.81 GPa based on the lamina elastic constants in Table 1. Lastly, to determine $g(\alpha)$, the dimensionless energy release rate, equation (10) is used in conjunction with a finite element analysis (FEA).

$$g(\alpha) = \frac{G(\alpha)E}{D\sigma_N^2} = \frac{JE}{D\sigma_N^2} \tag{10}$$

To provide accurate FEA results the J-integral method [29] and quarter element technique [30] is employed in Abaqus Standard-Implicit 2020. Multiple simulations are run at differing $\alpha$ values to obtain g as a function of $\alpha$ and a polynomial is then fit to these FEA results and its derivative evaluated to find $g'(\alpha)$. The resulting $g(\alpha)$ and $g'(\alpha)$ plots from this analysis are shown in Figure 13, $\alpha_o = 0.5$ in all the currently presented Gap Test data which corresponds to $g(\alpha) = 6.7176$ and $g'(\alpha) = 41.352$.

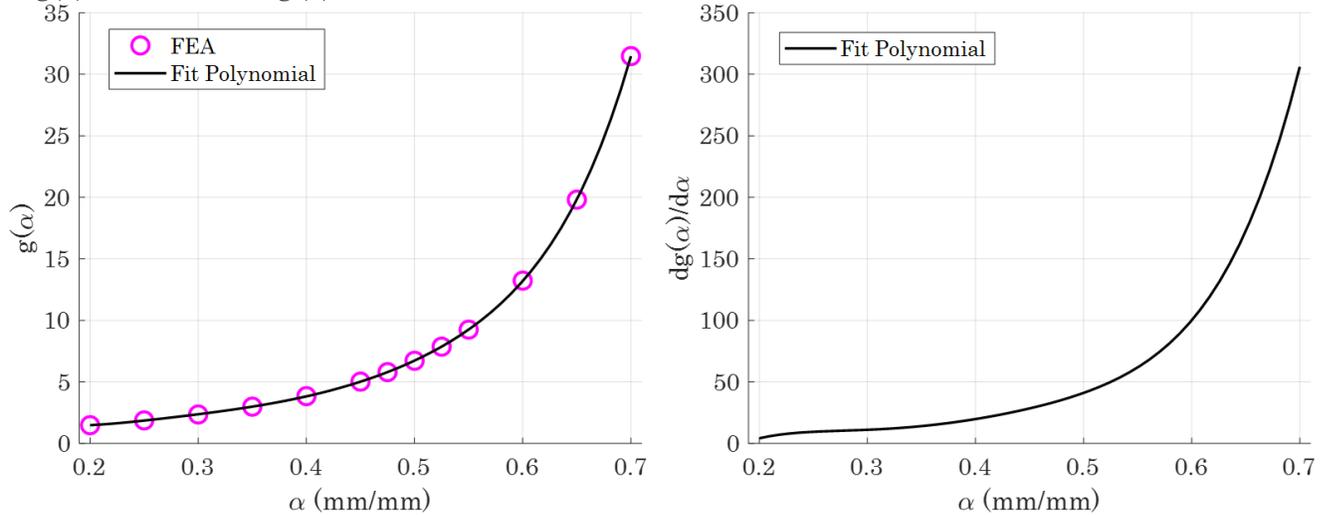

Figure 13 FEA simulation results for (left) the dimensionless energy release rate $g(\alpha)$ and (right) its derivative $g'(\alpha)$ as a function of the normalized crack length $\alpha = a/D$.

Using equation (6) leads to the combined linear regression plot for each $\xi$ data set in Figure 14, from such it is seen that the slope increases as $\xi$ increases which corresponds to a decrease in the fracture energy per equation (7). Similarly, the intercepts for $\xi = 0.29$ and 0.44 are larger than the intercept for $\xi = 0$ which indicates an increase in the FPZ size ($c_f$ value) due to the presence of a crack parallel stress. The decrease in fracture energy and increase in FPZ size are likely indicative of changing or increasing damage mechanisms at the crack tip that ultimately initiates fracture. Regardless, the analysis captured herein supports the hypothesis that the non-negligible FPZ in composites impacts the fracture behavior and a crack parallel compression leads to a weakening effect via reduction of the fracture energy. Numerical values for $G_f$ and $c_f$ from equations (7) and (8) are in Table 5.

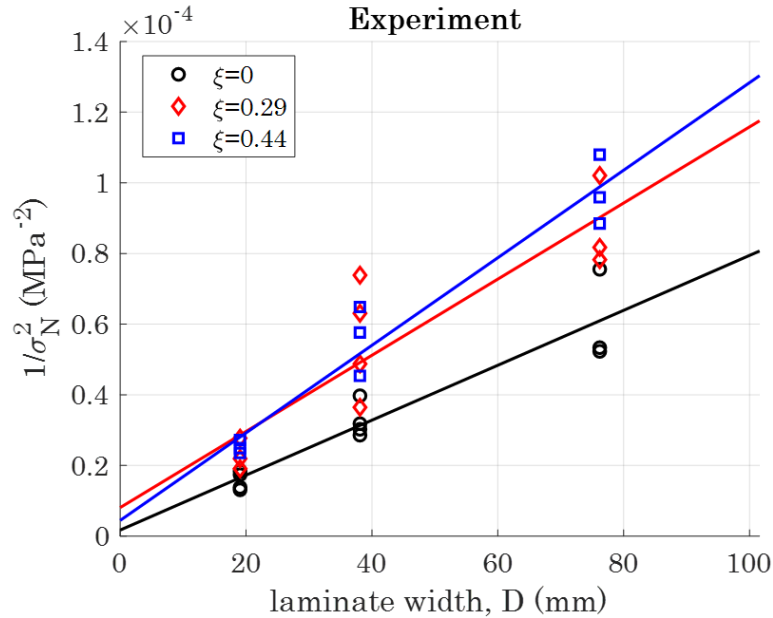

Figure 14 SEL linear regression plots for differing experimental $\xi$ values.

| $\xi$ | $G_f$ (N/mm) | % diff. | $c_f$ (mm) | % diff. |
|---|---|---|---|---|
| 0 | 102.85 | - | 0.361 | - |
| 0.29 | 74.20 | -27.86 | 1.229 | 240.44 |
| 0.44 | 64.52 | -37.27 | 0.587 | 62.60 |

Table 5 Calculated fracture energy $G_f$ and FPZ size $c_f$ for the experimental Gap Test data using Bažant's Type II Size Effect Law.

### 3.2 Crack Tip Fractography

Due to the foregoing results the failure mechanism(s) for each of the $\xi$ = 0, 0.29, and 0.44 tests are of keen interest. While the Bažant SEL provides insight to the structure's response under a crack parallel compression at the macro-scale (i.e., laminate level) the peculiarities of the composite's behavior at the meso-scale (i.e., lamina level) and micro-scale (i.e., matrix and fiber level) are unknown. In the present study, an optical microscope at high magnifications is used to obtain detailed images of the composite morphology at the crack tip where the fracture initiates. Photomicrographs for a $\xi$ = 0 test are shown in Figure 15 while that of a $\xi$ > 0 test is shown in Figure 16.

The $\xi$ = 0 test in Figure 15 is compared against a polished, pristine cross section which shows the mesoscale structure that is, the 0° and 90° layers and their respective interfaces are easily identifiable. When viewing the fractured surface, for the 90° plies (plies in the x direction) the fibers appear on a flat and smooth surface that is the fracture plane, implying minimal fiber breakage and instead a failure in the matrix while the 0° plies (plies in the y direction) have fractured completely, and fiber breakage is the observed failure mechanism.

This is indicative of a tensile dominant fracture which is expected due to the normal stresses from the globally applied bending moment. A limitation to this analysis though, is the absence of a timescale. Whether the fracture initiated via fiber breakage in the 0° plies or matrix cracking in the 90° plies first is unknown, and additionally the interaction of these two phenomena is undefined. Conventional wisdom suggests that matrix cracking in the 0° plies due to the tensile stresses would occur first, supported by the strength parameters of Table 2 where $X_T / Y_T = 2089.0 / 79.29 = \sim 26.3$. Though this is not a conclusion that may be made solely based on Figure 15. Nonetheless, fiber breakage in the 0° plies and matrix failure in the 90° plies are observed in all the $\xi = 0$ tests.

  The fracture surfaces for a $\xi > 0$ test, Figure 16, tell partially the same story as the fracture surface for a $\xi = 0$ test. Similar fiber breakage in the 0° plies and matrix failure in the 90° plies are present for the crack parallel compression tests, indicating that tensile failure is one of the mechanisms that drives crack propagation. But a considerably sized splitting crack is also observed for all of the $\xi = 0.29$ and $\xi = 0.44$ tests. Observationally the anatomy of these splitting cracks is always the same: one single splitting crack is present for each $\xi = 0.29$ or $0.44$ test, the splitting cracks only exist in the 0° plies, and the splitting crack is close to the laminate midplane though never exactly at the midplane. Additionally, the width of the resulting splitting cracks is higher in the $\xi = 0.44$ specimens than it is in the $\xi = 0.29$ specimens. Inspection of the micrographs for each $\xi$ level confirms that the presence of a crack parallel compression leads to a change in the fracture morphology which offers a possible explanation for the reduction in fracture energy that is shown in the previous sections. Via the initiation of a new and or additional failure mechanism it is likely that the crack propagates sooner and hence the load bearing capacity of the structure, and concomitantly the fracture energy, is reduced.

  However, this finding also complicates the present effort of diagnosing the crack initiation mechanism for $\xi = 0.29$ and $\xi = 0.44$. Previous researchers [31] performed 3PB tests on notched and un-notched composite specimens and showed that splitting cracks initiate mainly at the fiber/matrix interface, but also in the matrix, due to the high dilatational energy density at the interface. This damage then triggers localized crack propagation and ultimately catastrophic failure of the structure [32]. Unfortunately, though, due to the complex morphology at the crack tip of the Gap Test specimens these results from the literature are insufficient to uniquely define the crack initiation mechanism. In Figure 16 three different failure mechanisms are present – fiber breakage, matrix failure, and splitting cracks. And as before, the absence of a timescale in this analysis precludes one from isolating the governing failure mechanism from this list. To answer this question the Gap Test may be ran up to a percentage of the known peak load (i.e., $0.9P_{max}$), unloaded, and then the fracture surface analyzed as was done above. This will allow the identification of the first failure mechanism that manifests at the crack tip and consequently what micro- and meso-scale behavior dominates the structural strength in the presence of a crack parallel compression. The execution of this effort is beyond the scope of the present work and it is left for future studies.

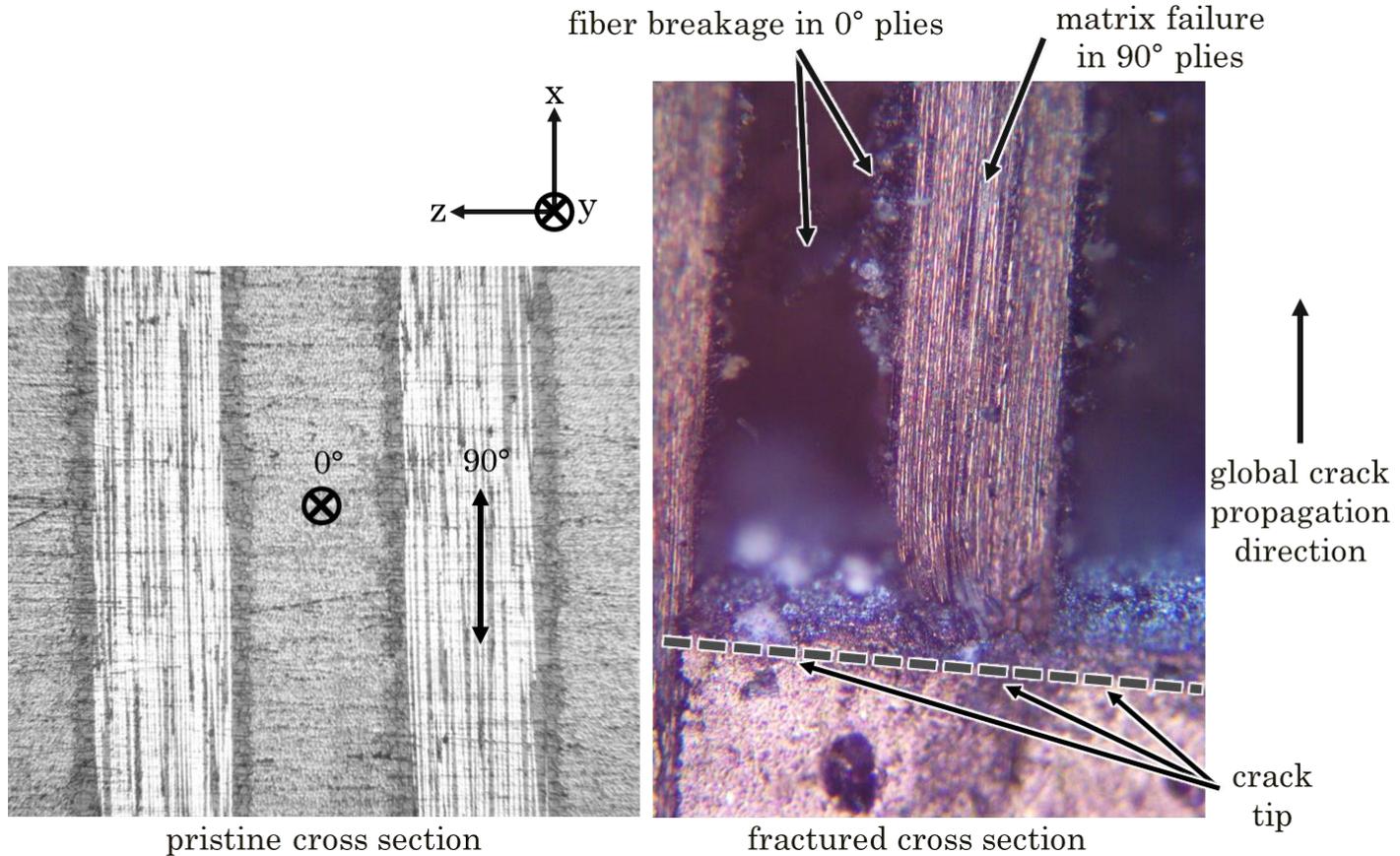

Figure 15 Representative micrograph of the crack tip morphology in a $\xi = 0$ test where fiber breakage in the 0° plies and matrix failure in 90° plies is observed. The fracture surface is compared against a pristine cross section which shows the mesoscale morphology prior to fracture.

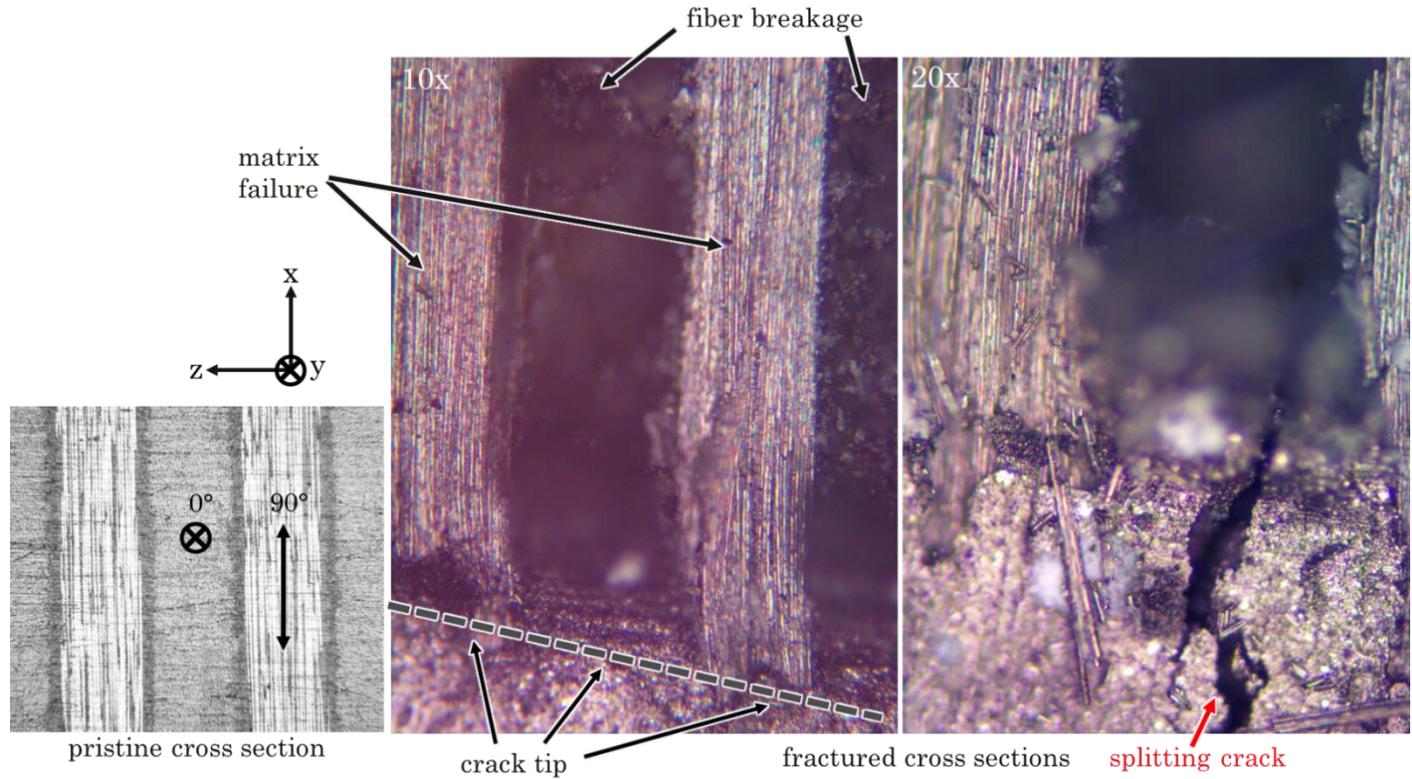

Figure 16 Representative micrograph of the crack tip morphology in a $\xi = 0.29$ test. The observed fracture surface displays similar fiber breakage and matrix failure to the $\xi = 0$ tests although, a significant splitting crack is now present due to the crack parallel compression.

## *3.3 Digital Image Correlation*

Digital Image Correlation (DIC) is also used to augment the load-line displacement that is output from the load frame transducer, which is a one-dimensional point value. DIC illustrates the displacement and strain fields on the surface of the test specimen and most critically, the development of the strain field adjacent to the crack tip. This is immensely beneficial as it elucidates the crack tip stresses while also being a surface representation of the FPZ and its evolution. DIC images showing the progression of the maximum principal strain for a $\xi = 0$ test are captured in Figure 17 while similar images for a $\xi > 0$ test are in Figure 18. Both for tests with and without crack parallel compression the highest intensity strain field originates at the crack tip and proceeds to increase in size as the applied load increases, an expected phenomenon due to the stress concentrations that develop at the crack tip [33]. For $\xi > 0$ tests the size of the intensified strain field at fracture is generally less than that in a $\xi = 0$ test, an observation that supports the preceding results of a reduction in fracture energy as $\xi$ increases. This is due to the additional damage that is induced within the FPZ from the crack parallel compression, as shown in Figure 16, which in turn reduces the structure's capacity to resist fracture and therefore the magnitude of the strain field (and consequently stress field) is lesser at failure.

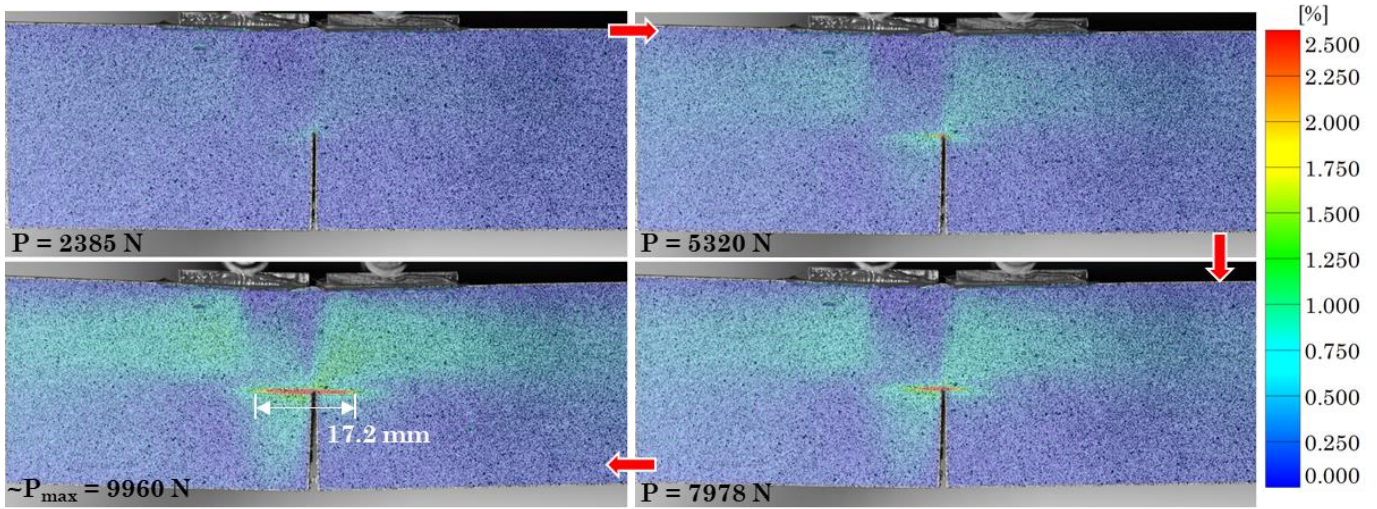

Figure 17 Maximum principal strain field evolution in a ξ = 0 test on a medium size specimen. Shown strain field is representative of all ξ = 0 tests.

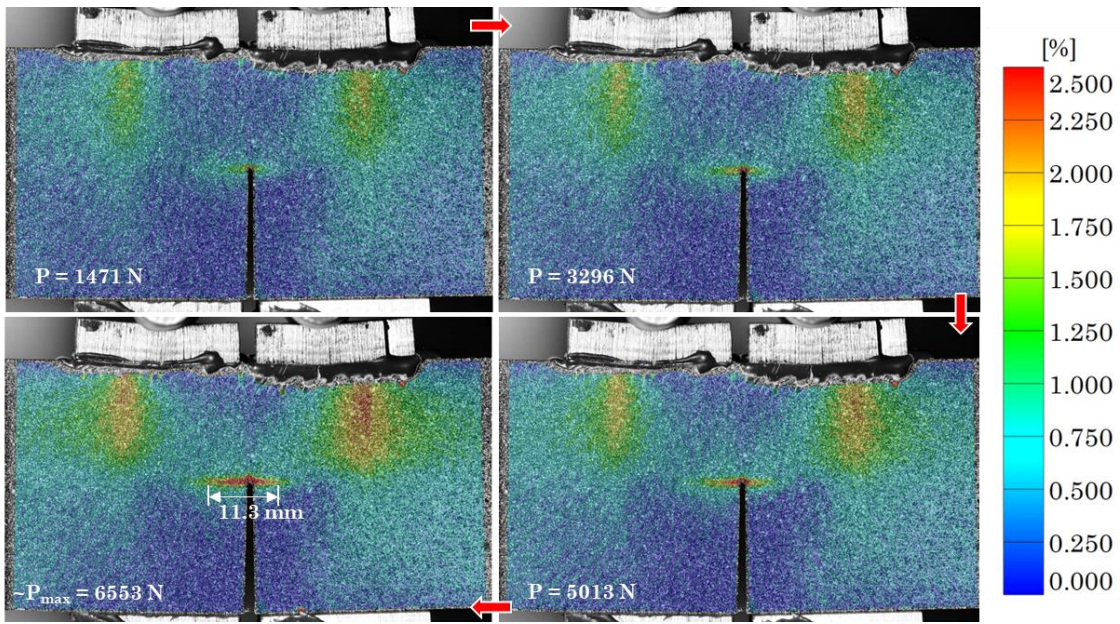

Figure 18 Maximum principal strain field evolution in a ξ = 0.29 test on a medium size specimen. Shown strain field is representative of all ξ > 0 tests

## 4 Computational Results and Analysis

The foregoing experimental campaign is buttressed with a thorough computational campaign to provide a framework that can capture the crack parallel compression effects observed in the Gap Test. Such an effort is tremendously important due to the current prevalence of composite materials in industry whose strength may be dangerously weaker than expected in the presence of a crack parallel compression. To this end, a computational framework allows the practicing engineer to appropriately model and predict a composite structure's behavior when crack parallel compression is experienced. Furthermore, correct modelling definition obviates the need to undertake a large experimental effort as done in this study.

## *4.1 Model Definition*

To corroborate the experimental Gap Test results a crack band modeling approach [34] is employed via Abaqus-Explicit 2020 simulations. The crack band model allows the inclusion of all the stress components in a material which, due to the presence of a finite width FPZ, plays a major role in fracture initiation of quasibrittle media [22, 23]. As has been previously stated traditional fracture theory and modeling approaches consider cracks as a line that is perfectly sharp. If such an approach is used currently it will be impossible to capture the influence of a crack parallel compression on fracture. To support this claim two different failure modeling techniques are leveraged – a Hashin damage law which is fully tensorial and a cohesive element model with a reduced tensorial damage law. These two approaches and their results are expounded upon in the following sections.

To computationally repeat the Gap Tests each specimen size listed in Table 3 is modeled with geometry, boundary conditions, and loading that replicates the experimental conditions. The composite laminates are modeled with the material properties in Table 1 and Table 2 and meshed with linear reduced integration continuum shell elements S4R to allow use of the Hashin damage law that is built into Abaqus. For $\xi > 0$ models the polypropylene pads are modeled with elastic and plastic stress-strain data, which is shown in Appendix B, acquired via compression tests per ASTM D695 [35] and meshed with linear reduced integration brick elements C3D8R. The elements used within the crack band are dependent upon the specific damage law adopted, these details are shared next.

### *4.1.1 Crack Band with Hashin Damage*

The Hashin damage law [36] predicts failure in a transversely isotropic material according to equations (11)-(14), which distinguish between and predict specific damage initiation mechanisms.

$$\textbf{Fiber Tension}: \left(\frac{\sigma_{11}}{X_T}\right)^2 + \left(\frac{\sigma_{12}}{S}\right)^2 = 1 \qquad (11)$$

$$\textbf{Fiber Compression}: \left(\frac{\sigma_{11}}{X_C}\right)^2 = 1 \qquad (12)$$

$$\textbf{Matrix Tension}: \left(\frac{\sigma_{22}}{Y_T}\right)^2 + \left(\frac{\sigma_{12}}{S}\right)^2 = 1 \qquad (13)$$

$$\textbf{Matrix Compression}: \left(\frac{\sigma_{22}}{2S}\right)^2 + \left[\left(\frac{Y_C}{2S}\right)^2 - 1\right]\frac{\sigma_{22}}{Y_C} + \left(\frac{\sigma_{12}}{S}\right)^2 = 1 \qquad (14)$$

Note that these equations are predicated upon a plane stress assumption which drives the general 3D stress tensor to a planar 2 by 2 tensor as shown in Figure 19. Because the Hashin criteria introduced above considers all these stress components within its failure criteria it is deemed a fully tensorial damage law. It is worth mentioning here that more sophisticated models for composites such as e.g. the spectral stiffness microplane model [37, 38] or the microplane triad model [39] could have been used. However, considering the simple layup investigated in this work, Hashin criterion was deemed sufficient to provide good estimates of the experimental results.

Now consider the stress state in front of the crack tip in the Gap Test which is biaxial due to the compressive and bending stresses, as shown in Figure 20. It is known that

19 of 30

quasibrittle materials exhibit a finite width FPZ and hence none of the in-plane dimensions of the crack, or effective crack that is the FPZ, are negligible. This implies that both the bending and compressive stresses will influence the crack propagation behavior. Further, if any fiber waviness or misalignment is present shear stresses will also develop in the FPZ due to the compression and bending stresses. Therefore all 3 stress components $\sigma_{11}$, $\sigma_{12}$, and $\sigma_{22}$ must be included in the failure criteria used when computationally modeling the Gap Test. Because equations (11)-(14) satisfy this requirement the Hashin damage criteria is used to predict failure in the crack band. This is achieved by simply meshing the crack band with identical elements that model the adjacent composite laminate, continuum shell S4R elements.

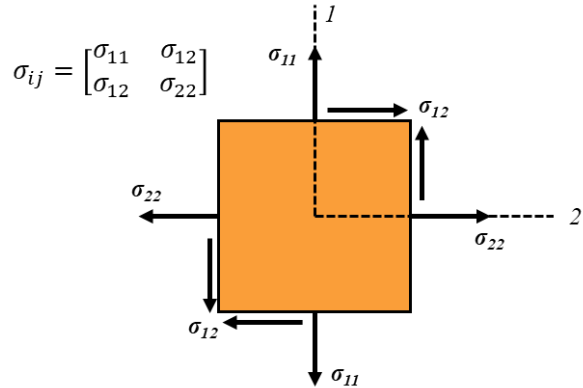

Figure 19 Reduced stress tensor for the case of plane stress used in the Hashin failure criteria in composites.

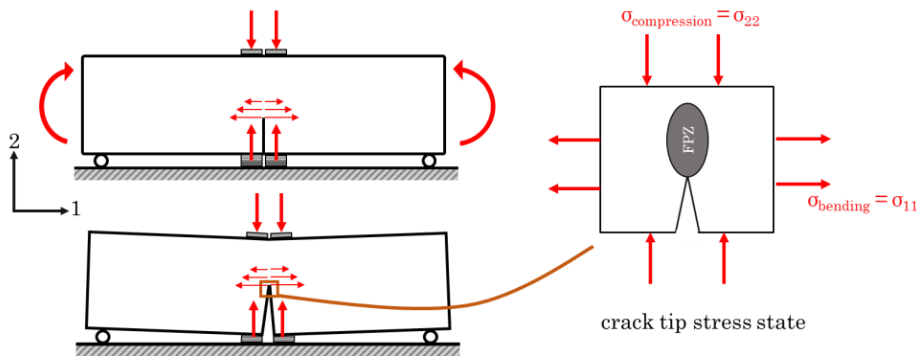

Figure 20 Crack tip stresses experienced during the Gap Test.

### *4.1.2 Cohesive Zone Modeling*

To further support the above reasoning the fracture is also modeled with a simpler, reduced tensorial damage law that is incapable of considering all the experienced crack tip stresses $\sigma_{11}$, $\sigma_{12}$, and $\sigma_{22}$. This serves to solidify two arguments, (1) a fully tensorial damage law is necessary to capture crack parallel compression effects and (2) the use of a reduced tensorial damage law can dangerously overpredict structural strength even when the crack band model is employed. This is accomplished by meshing the crack with cohesive elements COH3D8 [40] and a maximum stress failure criterion whilst leaving the remainder of the model identical to the previous simulations.

The constitutive relationship for a cohesive element is given via a linear elastic traction separation law that consists of one normal stress and two shear stresses, one in-plane and one out-of-plane [41]. These stress components are also uncoupled. Note that when this constitutive relationship is used to model the crack it may only experience stresses that coincide with the 3 traditional fracture modes defined in LEFM – $\sigma_n$ is an opening stress (Mode I), $\tau_s$ is in-plane shear (Mode II), and $\tau_t$ is out-of-plane shear (Mode III). This effectively reduces the crack model to a line crack model. More importantly though, fracture initiation occurs when any of the stress components $\sigma_n$, $\tau_s$, or $\tau_t$ reach a critical value. This is represented by the following equation where $\sigma_{n,crit}$, $\tau_{s,crit}$, and $\tau_{t,crit}$ refer to nominal material strengths for each mode of loading:

$$max\left\{\frac{|\sigma_n|}{\sigma_{n,crit}}, \frac{\tau_s}{\tau_{s,crit}}, \frac{\tau_t}{\tau_{t,crit}}\right\} = 1 \qquad (15)$$

Thus, the tensorial damage law used presently to define the cohesive elements in the crack can be written as follows:

$$\begin{Bmatrix} \sigma_n \\ \tau_s \\ \tau_t \end{Bmatrix} = (1-D)\begin{bmatrix} E_n & 0 & 0 \\ 0 & G_s & 0 \\ 0 & 0 & G_t \end{bmatrix}\begin{Bmatrix} \epsilon_n \\ \epsilon_s \\ \epsilon_t \end{Bmatrix} \qquad (16)$$

where ε denotes the strain component, $E$ and $G$ are elastic constants, and $D$ is a damage variable that monotonically increases from 0 to 1 upon further loading after damage has initiated per equation (15). As is quickly noticed, equation (16) is incapable of capturing a crack parallel compression due to the simplicity of the tensorial damage law. Thus, when cohesive elements are used with a reduced tensorial damage law in the crack band the experimental results observed in Table 5 are impossible to capture. This point is further illustrated in Figure 21.

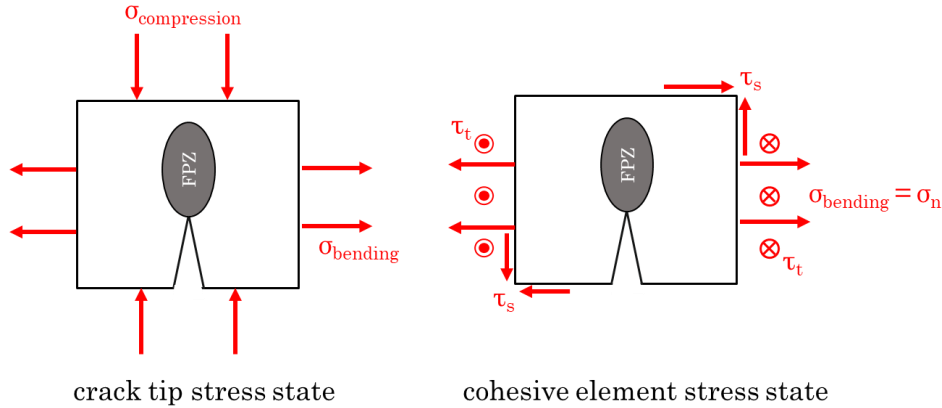

Figure 21 Comparative schematic of (left) the actual crack tip stresses during the Gap Test and (right) the stress state captured in a cohesive element. Notice that the cohesive element does not capture the effects of a crack parallel compression.

### *4.2 Simulation Results*

For the Hashin damage law all the needed material properties for the computational model are standard properties given in Table 1 and Table 2 and the experimentally determined fracture energy in Table 5. But for the cohesive element model there are 3 non-standard properties needed that are not readily available: $\sigma_{n,crit}$, $\tau_{s,crit}$, and $\tau_{t,crit}$ which denote the material strengths as defined in Figure 21 and effectively represent smeared strengths of the composite laminate. Hence, the Tsai-Wu failure criterion [28] is used to provide an initial guess for $\sigma_{n,crit}$, $\tau_{s,crit}$, and $\tau_{t,crit}$ and ξ = 0 simulations are run and checked against the ξ = 0 experimental data. The $\sigma_{n,crit}$, $\tau_{s,crit}$, and $\tau_{t,crit}$ values are then iteratively updated until the ξ =

0 simulations match the $\xi = 0$ experiments. The calibrated $\sigma_{n,crit}$, $\tau_{s,crit}$, and $\tau_{t,crit}$ values are given in Figure 22 with a comparative force-displacement plot of the $\xi = 0$ cohesive simulations and experiments, Figure 22 also includes the simulation results for shell elements and the Hashin damage law in the crack band.

Both computational models do a sufficient job of matching the stiffness, strength and post-peak response of the experimental $\xi = 0$ data. This is expected as for the case of no crack parallel compression both damage laws capture all the experienced crack tip stresses. It is worth noting that both the models match the post-peak very well for the small and medium size specimens. However, they both fail to predict the progressive failure behavior in the large specimens. The reason is that the softening law in composite materials has been shown to be bilinear [16]. However, in order to keep the number of parameters low and keep the analysis as clear as possible, a linear softening approximation was adopted in this investigation (note that most of the commercial software to simulate composites assumes a linear softening law). When the specimen size increases, the FPZ develops more and the cohesive stresses at the crack tip reach the second branch of the softening curve. These stresses contribute to energy dissipation and lead to a more progressive softening. However, the linear softening laws used in this project are not able to capture this aspect. While the authors could have adopted bilinear softening laws, this would have added two more parameters to the model: the total fracture energy $G_F$ and the stress corresponding to the change in slope $\sigma_k$. Such parameters would have been extremely cumbersome to characterize since more specimen sizes would have to be tested. Moreover, the presence of the additional parameters would have complicated the analysis of the results. Since the linear softening models capture the behavior up to the peak load very well and the focus of this study is not on the post-peak behavior but on the peak-load, the authors decided that a linear softening model represented a very good compromise.

Based on the accuracy of the $\xi = 0$ simulations corresponding simulations are ran for $\xi = 0.29$ and 0.44. The peak load values from these simulations are plotted in Figure 23 and tabulated in Table 6. As was foreseen, the fully tensorial Hashin damage law predicts a similar trend as the experimental data – as $\xi$ is increased the peak load (i.e., structural strength) decreases. This is possible due to the crack parallel compression being included in the damage initiation criterion of equations (11)-(14). Also as predicted, the reduced tensorial damage law implemented with cohesive elements is unable to capture the crack parallel compression effects and the reduction in structural strength is not captured. Per Table 6 it is seen that, for a compression of $\xi = 0.44$, the structural strength is overpredicted by the cohesive elements between 23.6% and 37.2% based on the panel size. Although not as severe, structural strength overpredictions are also seen for the lower $\xi$ level with cohesive elements.

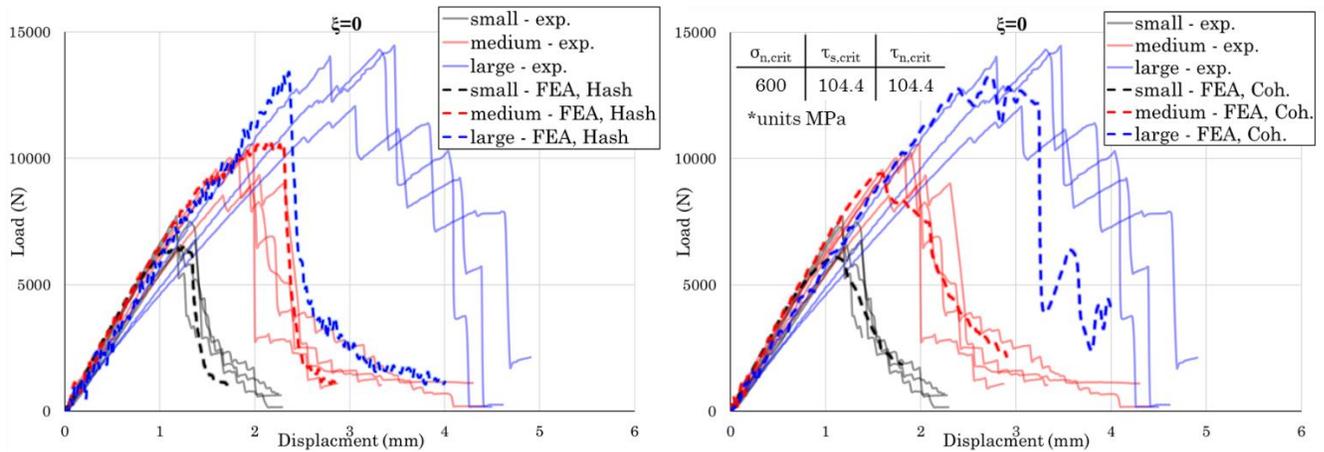

Figure 22 Experimental force-displacement data for ξ = 0 (solid lines) superimposed with simulated data (dashed lines) when (left) the Hashin damage models the crack band and (right) equivalently for when Cohesive elements model the crack band. Calibrated strength parameters are also shown.

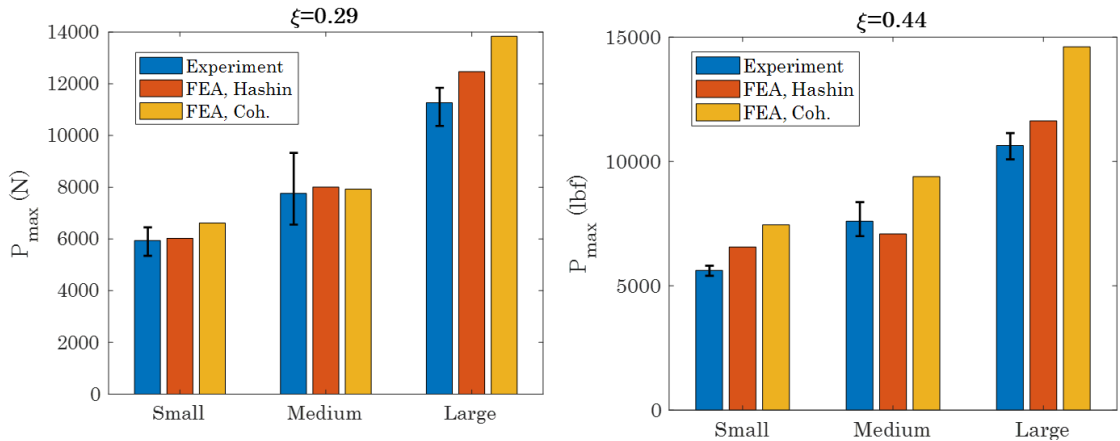

Figure 23 Measured peak loads for (left) ξ = 0.29 and (right) ξ = 0.44 compared against simulated peak loads.

|  |  | Simulated Peak Loads (N) | | | |
| --- | --- | --- | --- | --- | --- |
|  | Size | Hashin Damage | % diff. | Cohesive Elements | % diff. |
| ξ = 0.29 | Small | 6024.6 | 1.5 | 6614.4 | 11.4 |
|  | Medium | 8007.8 | 3.2 | 7925.8 | 2.1 |
|  | Large | 12467.5 | 10.6 | 13832.2 | 22.7 |
| ξ = 0.44 | Small | 6559.94 | 16.7 | 7450.8 | 32.6 |
|  | Medium | 7082.17 | -6.7 | 9389.1 | 23.6 |
|  | Large | 11631.4 | 9.3 | 14609.8 | 37.2 |

\* % difference is calculated from the experimental average of each ξ level

Table 6 Simulated peak load values for ξ = 0.29 and 0.44

# 5 Conclusion

This paper studied, for the first time, the effect of a crack parallel compression on the global Mode I fracture behavior of a carbon fiber polymeric matrix composite. To achieve this goal, an avant-garde experimental technique, originally coined elsewhere as *The Gap Test,* was designed, developed, and its nuances thoroughly documented for applicability to composites. To extend the experimental results, a computational effort was also undertaken that provides a framework capable of capturing the crack parallel compression effects. Based on these endeavors the following conclusions are made:

1. For notched cross ply composites the presence of a crack parallel compression leads to a reduction in the structural strength and fracture energy. Specifically, as the crack parallel compression is increased the structural strength and fracture energy monotonically decreases. This challenges the century old hypothesis that fracture energy is a constant material property and complicates the use of fracture energy as a single parameter description of fracture, which is used in many design settings. The observed weakening is explained in three separate but intertwined ways:
    (1) the Gap Test features two distinct, statically determinant loading regimes (compression and bending) that unequivocally shows a reduction in peak load when a crack parallel compression is present, i.e., the structures ability to resist fracture is diminished;
    (2) using photo-microscopy, a clear change in fracture morphology is induced when crack parallel compression precedes a bending, crack opening force. That is, splitting cracks develop at the crack tip which expedite catastrophic crack propagation. For no crack parallel compression splitting cracks are not present at the crack tip and instead fiber breakage and tensile matrix failure are the dominant failure mechanisms – which is the expected response;
    (3) leveraging digital image correlation to analyze the experimental results, the magnitude of the maximum principal strain field close the crack tip at peak is lesser for the case of crack parallel compression than for no crack parallel compression. Indicating that crack parallel compression causes failure at smaller strain intensities at the crack tip.
2. Composites exhibit a non-negligible fracture process zone that significantly affects their structural behavior in the presence of a crack/defect. This is phenomenological and hence the line crack assumptions that Linear Elastic Fracture Mechanics are founded upon are inappropriate to apply to composites. Quasibrittle Fracture Mechanics is necessary to predict a cracked composites response and specifically, the size effect scaling of structural strength that is inherent to quasibrittle materials must be accounted for.
3. The ability of a crack parallel compression to reduce structural strength indicates that cracks in a composite may open in a non-traditional mode that is not defined by the standard Mode I, II, or III loading. This is explained by the non-negligible fracture process zone which has finite dimensions that are normal to the crack parallel compression force vector.
4. To computationally capture the effects of crack parallel compression in composites a crack band modeling approach must be used and further, the crack band must be defined by a fully tensorial damage law. If a reduced tensorial damage law is used, that does not allow for the influence of a crack parallel compression, specious results will be obtained that overpredict the true structural strength.

5. The experimental configuration presented in this study is feasible for composites, but the authors have 2 main lessons learned to share:
    (1) the use of lateral stability columns is compulsory for experimental success when loading the thin composite test specimens in compression. Had this design choice been made initially, approximately 4 weeks of experimental development would have been avoided. Stability columns additionally simplify the buckling analysis and permit testing on thinner specimens that are easier to manufacture;
    (2) lower compressive loads generally lead to less cumbersome testing and analysis as the yield plateau is reached faster, densification of the polypropylene pads is avoided, and damage to the test fixture does not occur. In the Gap Test design phase, the lowest compressive loads that permit testing of the desired ξ values should be used.

## 6 Final Remark

The experiments defined herein demonstrate the criticality of crack parallel compression, particularly due to the observed weakening effect which would prove very dangerous in a real-world structure. But the results are too limited to immediately warrant drastic changes in composite fracture theory or tools used by practicing engineers. Different layup sequences, material systems, and geometries must be tested and further, the individual effects of crack parallel tension and shear, both in-plane and out-of-plane, need to be evaluated. The behavior under fatigue and mixed mode loading will also require attention, evaluated based on all the aforementioned conditions. Clearly, a substantial amount of resources, financing, and ambitious researchers are necessary to fully define the behaviors observed by the present study.

## Appendix A: Buckling Analysis

To satisfy equation (2) and generate the geometric values in Table 3 an analytical and computational approach is employed to predict the fracture and buckling loads for the largest test specimen size desired. Buckling is a phenomenon that increases in likelihood for structures with greater aspect ratios so, it is trivial to perform the same analysis for smaller specimens.

Fracture is predicted according to the LEFM expression:

$$\sigma_N = \sqrt{\frac{E G_f}{D g(\alpha_o)}} \tag{17}$$

Where structural strength $\sigma_N$ is defined per equation (9) which when combined with equation (17) gives the fracture load.

$$P_{fracture} = \frac{2D^2 t}{3(L-s)} \sqrt{\frac{E G_f}{D g(\alpha_o)}} \tag{18}$$

The buckling load is predicted using an eigenvalue buckling analysis that is built in to Abaqus-Implicit [42, 43]. This method functions via a linear perturbation technique that determines the loads for which the model stiffness matrix becomes singular.

Now, from a design perspective the dimensions a, L, D, and s in Figure 4 are iterated through parametrically while the thickness t is left as an independent variable. The buckling and fracture loads are then evaluated until a geometry that meets equation (2) is found. This results in the plot in Figure 24, which is based on the large panel dimensions in Table 3, where η = 2 is used as the fracture load is based on an LEFM analysis which is only loosely appropriate for composites.

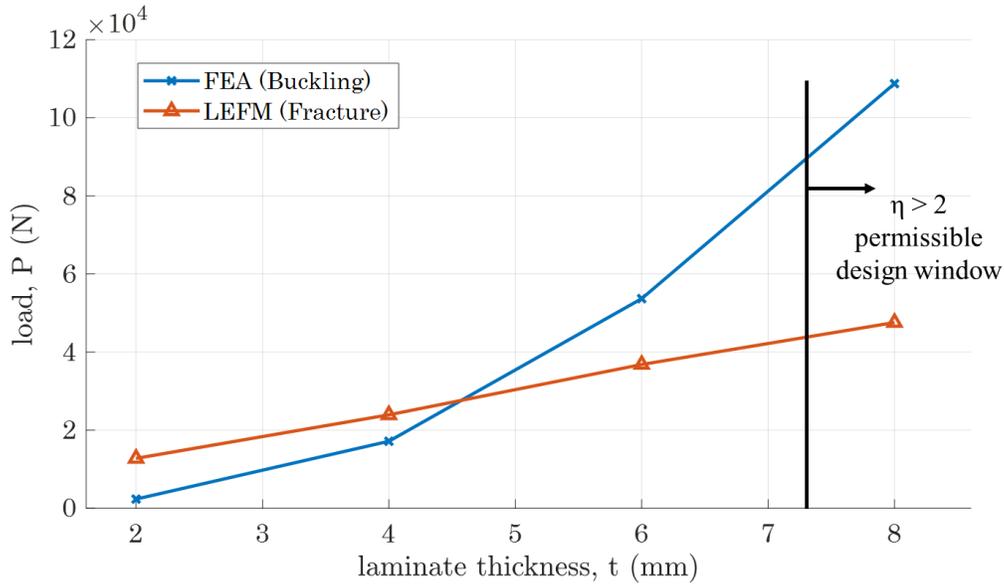

Figure 24 Predicted buckling versus facture loads with a factor of safety η = 2 applied.

## Appendix B: Polypropylene Stress-Strain Curves

Based on the efforts of the concrete Gap Tests [22, 23] it is known that polypropylene exhibits a near perfectly plastic yield plateau but, the plateau stress and strain must be known to use the polypropylene in a design sense. In accordance with ASTM D695 [35] ten polypropylene samples are tested to generate the following stress-strain curves.

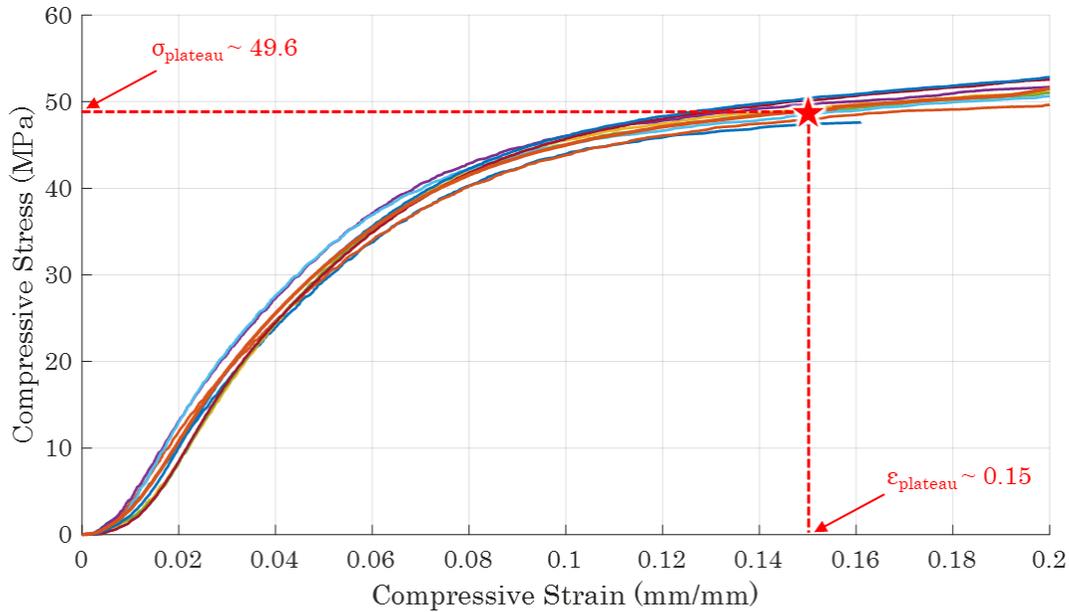

Figure 25 Compressive stress-strain curve of polypropylene showing the determined plateau stress and strain. Data is obtained from the ASTM D695 test procedure.